\documentclass[
  aip,
  rsi,
reprint,
  amsmath, amssymb,
  ]{revtex4-2}
\usepackage[utf8]{inputenc}
\usepackage[T1]{fontenc}
\usepackage{mathptmx}
\usepackage[
]{xcolor}
\definecolor{linkblue}{RGB}{31,119,180}
\usepackage[
  unicode,
  colorlinks,
  citecolor=blue,
  linkcolor=linkblue,
  urlcolor=linkblue,
  bookmarks=true,
  bookmarksopen=true,
  bookmarksopenlevel=3,
  bookmarksnumbered=true]{hyperref}
\usepackage{siunitx}
\usepackage{graphicx}
\usepackage{amsmath}
\usepackage{amssymb}
\usepackage{textcomp}
\usepackage{listings}
\usepackage{courier}
\usepackage{longtable}
\usepackage{booktabs}
\usepackage{pgf}

\newcommand{\figref}[1]{figure~\ref{fig:#1}}

\def\nqontrol{\texttt{NQontrol}}
\def\samplefreq{\SI{200}{kHz}}
\def\simulchans{eight}
\def\numsos{five}
\def\totalsos{40}
\def\eg{e.g.\ }
\begin{document}

\preprint{AIP/123-QED}

\title{NQontrol: An open-source platform for digital control-loops in quantum-optical experiments}
\author{Christian Darsow-Fromm}
    \email{christian.darsow@physik.uni-hamburg.de}
\author{Luis Dekant}
\author{Stephan Grebien}
\author{Maik Schröder}
\author{Roman Schnabel}
    \affiliation{Institut für Laserphysik und Zentrum für Optische Quantentechnologien der Universität Hamburg, Luruper Chaussee 149, 22761 Hamburg, Germany}
\author{Sebastian Steinlechner}
    \email{s.steinlechner@maastrichtuniversity.nl}
    \affiliation{Department of Gravitational Waves and Fundamental Physics, Maastricht University, P.O. Box 616, 6200 MD Maastricht, The Netherlands}
    \affiliation{Nikhef, Science Park 105, 1098 XG Amsterdam, The Netherlands}

\begin{abstract}
    Here we present NQontrol, a digital feedback-control solution based on the ADwin platform that delivers \simulchans{} simultaneous feedback loops running with \samplefreq{} sampling frequency, and offers five second-order filtering sections per channel for flexible shaping of the feedback loop. With this system, we demonstrate a Pound-Drever-Hall lock of an optical resonator and compare its performance to an analog reference implementation.
A comprehensive support package written in Python, together with a web-based graphical user interface (GUI), makes the system quick to setup and easy to use, while maintaining the full flexibility of open-source platforms.
 \end{abstract}

\maketitle

\section{Introduction}
\label{sec:introduction}

Control loops are a fundamental part of many experiments in quantum optics.
They are used to precisely control (``lock'') the phase relation of laser beams, keep optical cavities on resonance, stabilize lasers to atomic transitions, and much more \cite{bechhoefer2005,abramovici_feedback_2000}.
Depending on the subject area, a clustering of different hardware implementations for these control loops can be observed, with designs and approaches shared when researchers move between groups.
Some groups -- including ours until recently -- solely rely on self-built analog electronics.
Others have successfully implemented control loops with microcontrollers \cite{huang_microcontroller-based_2014} or FPGA boards \cite{neuhaus_pyrpl_2017}.
Working groups close to large collaborations, \eg particle physics or gravitational-wave astronomy, tend to use the purpose-built control and data acquisition systems of these fields \cite{epics,desy_doocs_nodate,bork_advligo_nodate}.
Commercial solutions are also available and successfully used in some applications.

Each of these approaches has its own advantages and limitations.
Analog control loops allow very high control bandwidths (many MHz) with very low noise, as they do not suffer from digital quantization issues.
However, they are time-consuming to build and change, so convergence to an optimal controller design is slow.
Dynamical adjustment of filters, as well as automation and interfacing between several control loops is difficult to do.
Microcontroller based circuits can be cheap solutions, and a wealth of development tools and add-ons exists \eg in the Arduino landscape \cite{arduino}. On the other hand, these micro controllers usually cannot reach high control bandwidths and their built-in analog-to-digital conversion is of low resolution or poor noise performance.

FPGAs overcome these limitations when they are interfaced to fast, high-accuracy analog-digital converters (ADCs).
Designing and building suitable circuit boards for these complex chips is, however, very involved: FPGAs come in high-density packaging that require carefully matched signal delays and reflow soldering capabilities on multi-layered boards.
In addition, FPGAs generally have to be programmed in a hardware-description language (HDL), which is significantly different from general programming languages and poses a high barrier for development with FPGAs in small research groups.
A few years ago, small FPGA boards including ADC converters became available for an affordable price tag, \eg the RedPitaya/STEMlab \cite{redpitaya} family.
These boards, if they fit the requirements of the control task, reduce the problem of having to design own circuit boards to having to develop suitable interfaces to the experiment.
Programming such boards is still rather involved, but the \texttt{pyrpl} project \cite{neuhaus_pyrpl_2017} has developed a sophisticated control and analysis software package, which might satisfy common control tasks.

Large-scale systems tend to require significant investments in terms of hardware and trained personnel to get set up, putting them out of reach of most research groups.
Lastly, commercial ready-made controllers have the disadvantage that they are often tied to a specific application, or have only very basic PID (proportional/integral/differential) functionality.
In many cases, much better control performance could be achieved by custom-tailoring filter functions, \eg with second-order filters.

Here, we present \nqontrol \cite{nqcontrol}, a control solution based on the ADwin \cite{adwin} platform, which is a modular data acquisition and control platform consisting of a real-time computing unit and several input/output modules.
Our implementation can handle up to \simulchans\ simultaneously running control loops at a sample rate of \samplefreq, each having an arbitrarily defined filter function of up to 10th order (five second-order sections).
We provide an easy-to-use software interface written in \texttt{Python}, together with a web-based graphical user interface.
The real-time code is written in the ADwin \texttt{BASIC} dialect, which compiles quickly and provides a well-documented interface to the input/output modules.

This paper is laid out as follows. 
We first list our design considerations for the control system, establishing the use-cases and requirements that we set for our digital control system. 
Afterwards, we shortly introduce the structure of our system and review the basics of second-order filters and their implementation in digital systems. 
An example usage of the user interface is also given. 
Finally, we compare the achieved performance of our system in an experimental setup with a traditional analog controller, each optimized for the same control task of about \SI{4.5}{kHz} bandwidth. 
We find that our new system gives comparable performance, with the large advantage of providing much higher flexibility and being able to change control parameters at the click of a button.
 \section{Design Considerations}
\label{sec:designconsiderations}

Several considerations have influenced the design of our control system, which can be summarized as follows:

\begin{enumerate}
    \setlength\itemsep{0pt}
    \item It should be based on an \emph{established hardware platform}, commercially available and long-term supported, such that new systems are easy to obtain and set up for years to come.
    \item A wide input voltage range, with differential sensing and standard connectors, should \emph{not introduce new external interfacing} electronics.
    \item Ideally, the system should \emph{simultaneously support all locking loops} of an experiment, or be sufficiently modular to support those.
    \item A \emph{control bandwidth} (unity-gain point) of at least \SI{10}{kHz} should be realizable, with \emph{low electronic noise} and \emph{high resolution} to be able to compete with analog designs.
    \item The system should \emph{operate in real time}, for deterministic behavior of the control loops with an amplitude-phase response that does not depend on system load.
    \item Hardware \emph{programming should be accessible} with little more than the standard programming training that is expected of students in the
    STEM fields.
    \item Control loop filtering should go beyond PID control, allowing for \emph{arbitrarily defined biquadratic filters}.
    \item There should be an \emph{easy-to-use remote-control interface}, allowing quick results for daily lab work, as well as being accessible for continued development.
\end{enumerate}

Based on items (1)-(5), we decided that a modular system with a dedicated, real-time processor would best fit our requirements.
Such systems are \eg based on the PXI platform, or vendor-specific implementations.
Because of existing experience in our institute, we settled on the ADwin \cite{adwin} platform.
This platform also mostly fulfills (6), as it is programmed in a relatively easy to use BASIC dialect and the development environment is easily set-up and well documented.
Items (7) \& (8) then are the design considerations for the software that we developed for this platform, and that we will further describe in the following sections.
 \section{Implementation}
\label{sec:implementation}

\begin{figure}
    \includegraphics[width=\linewidth]{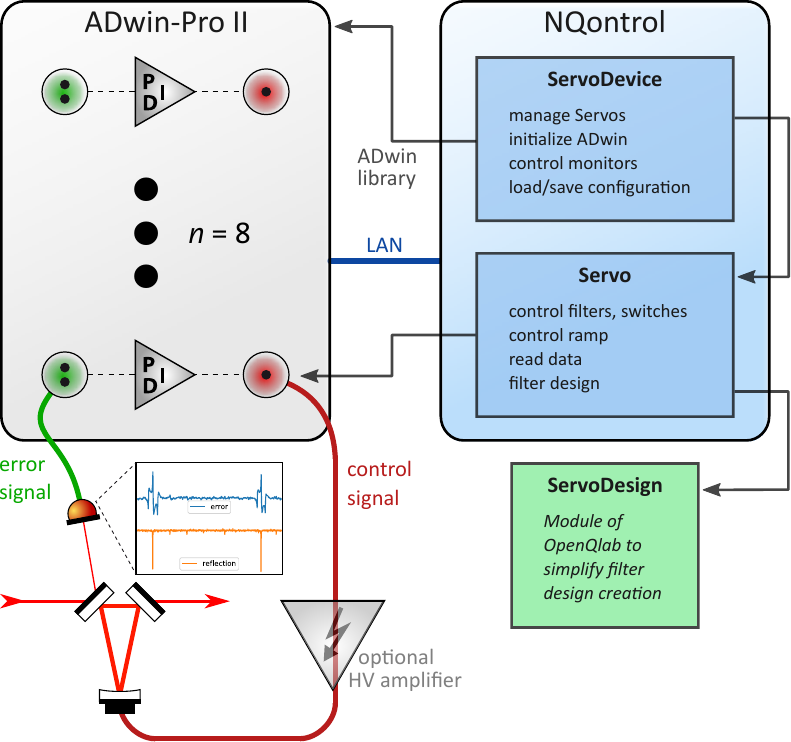}
\caption{
    System overview to show the relations between the experimental control loop, the real-time system (ADwin-Pro II) and our software package.
    The real-time system runs independently and is accessible via local network.
    \nqontrol\ changes the state of the real-time control loops and provides an interface for the user.
    It mainly consists of the classes \texttt{ServoDevice} that represents the whole real-time device and \texttt{Servo} which represents one of the \simulchans\ controllers.
  }
  \label{fig:system_overview}
\end{figure}

\subsection{System Description}
\label{sec:system}
We use an ADwin-Pro II system with a \SI{1}{GHz} ARM processor, $2\times 8$ simultaneously sampling 16-bit analog inputs and $2\times 8$ 16-bit analog outputs. The system itself runs some variant of the Linux operating system, which is however inaccessible to the end user, and handles the ADwin-proprietary communication with the hardware modules. It also provides a shared memory region that is accessible to a connected computer through a gigabit Ethernet connection for fast exchange of data. On top of this software platform, our high-priority real-time control code runs with a fixed cycle frequency, which we chose to be \samplefreq.

\begin{figure}
    \includegraphics[width=\linewidth]{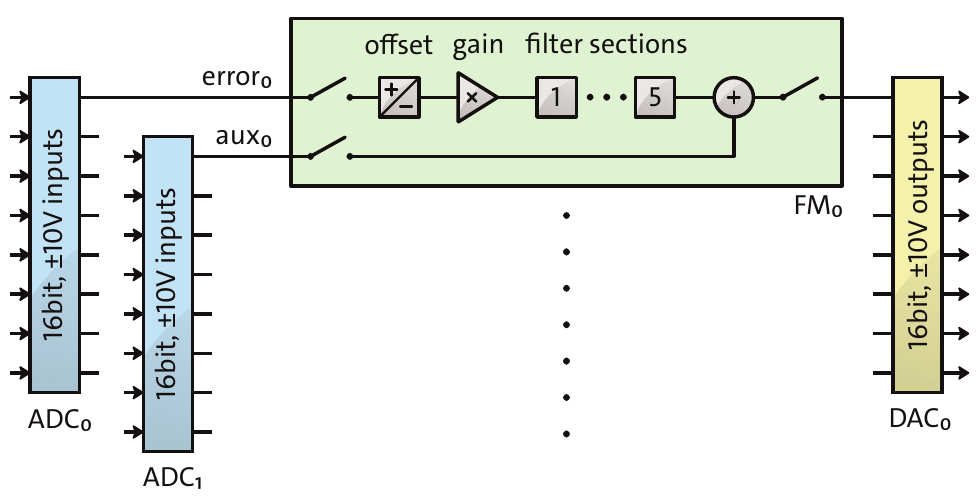}
    \caption{
        Block diagram overview of the individual channels in our digital control setup. 
        Each of the \simulchans\ channels provides offset correction, gain and a filter section.
        Additionally, an auxiliary signal can be added to the output, e.g.\ for frequency-response analyses.
        The input and output channels have a resolution of \SI{16}{bit}, where the voltage range of the input channels can be adjusted at four steps from $\pm\SI{1.25}{V}$ to $\pm\SI{10}{V}$.
    }
    \label{fig:io_design}
\end{figure}

Our controller has \simulchans{} independent control channels, comprised of a filter module $\text{FM}_i$ running as part of our real-time application.
The filter modules each take two analog input channels and provide one analog output channel, as depicted in \figref{io_design}.
One input channel is used for the error signal, while the other serves as an auxiliary signal that can be used for \eg response function measurements, monitoring or as a trigger input for lock automation.

On the software side, the filter modules have enable switches for the inputs and output and offer offset and gain correction.
They run with double-precision (\SI{64}{bit}) floating point arithmetic, to reduce rounding errors during the filter calculation.
The actual filter response is given by five second-order sections, described below, that can be individually activated.

A user-selectable combination of signals can be sent to four additional analog outputs, such as a copy of the input, auxiliary and output channels, to allow for easy monitoring.

\subsection{Second-Order Sections}
\label{sec:sos}

Digital systems generally run with a fixed sampling rate $f_s$.
Thus, there exist discrete sequences $x[i]$ and $y[i]$ that describe the input and output values, respectively, of the system at times $t_i$.
The entries of those sequences are spaced at time-intervals $T$ corresponding to the inverse sampling rate, $T = f_s^{-1}$.
A \emph{linear time-invariant discrete} filter acts on these histories of previous inputs and previously calculated outputs to produce a new filter output $y[n]$ \cite{bechhoefer2005},
\begin{align}
  y[n] = \sum_{i=0}^N b_i x[n-i] - \sum_{i=1}^M a_iy[n-i]\,.
  \label{eq:diff_eq}
\end{align}
In practice, and because of the feedback-nature of the $a_i$ coefficients, only a handful of these coefficients will be non-zero.
The order of the filter is given by the greater of $N$ or $M$.

Similarly to the Laplace transform in analog filter design, which converts time-domain signals into a frequency-domain analysis, digital filters use the $z$-transform
\begin{align}
  X(z) = \mathcal{Z}(x[n])=\sum_{n=-\infty}^\infty x[n]\,z^{-n}\,,
\end{align}
where $z = e^{sT}$ and $s$ is a complex number.
In particular, calculating the transfer function $H(z)$ of a discrete system given its inputs $X(z)$ and outputs $Y(z)$ is straight forward:
\begin{align}
  H(z) = \frac{Y(z)}{X(z)}\,.
\end{align}
$H(z)$ can be expressed as the quotient of two polynomials with coefficients $a_i$, $b_i$ from Eq.~\eqref{eq:diff_eq},
\begin{align}
  H(z) = \frac{\sum_{n=0}^N b_n z^{-n}}{1 + \sum_{m=1}^M a_m z^{-m}}\,.
\end{align}
The frequency response of such a system in the $z$-domain can be obtained by evaluating it on the unity circle $z = e^{i\omega T}$, where $\omega$ runs from $-\pi/T$ to $+\pi/T$.

Because the order of magnitude of the polynomial coefficients tends to diverge quickly, high-order discrete filters can run into numerical inaccuracies.
This is mostly resolved by breaking high-order filters down into consecutive sections of second order, i.e.\ of the following form (normalized such that $a_0 = 1$):
\begin{align}
  H(z) = \frac{b_0 + b_1 z^{-1} + b_2 z^{-2}}{1 + a_1 z^{-1} + a_2 z^{-2}}\,.
  \label{eq:sos}
\end{align}
Putting these coefficients into Eq.~\eqref{eq:diff_eq}, the time-domain equation of a second-order section is given by
\begin{multline}
  y[n] = b_0x[n] + b_1x[n-1] + b_2x[n-2] \\
    - a_1y[n-1] - a_2y[n-2]\,.
\end{multline}
Our real-time code then employs the so-called direct form II for calculating the filter,
\begin{multline}
  y[n] = c_0\bigl(x[n] - (c_1 + c_3)w[n-1] \\
    - (c_1 + c_4)w[n-2]\bigr)\,,
  \label{eq:df2}
\end{multline}
using the five double-precision coefficients $c_0 = b_0$, $c_1 = a_1$, $c_2 = a_2$, $c_3 = b_1/b_0$, and $c_4 = b_2/b_0$, as well as two \emph{history} variables $w[n-1]$ and $w[n-2]$, given by
\begin{align}
  w[n] = x[n] - c_1w[n-1] - c_2w[n-2]\,.
\end{align}

\subsection{Software}
\label{sec:software}

Our software implementation consists of two parts: a high priority real-time process running on the ADwin device and a platform-independent Python program which controls the real-time process via a network connection, controlling its state and providing filter coefficients (see figure~\ref{fig:system_overview}).
Using this combination it is easy to change and optimize the control parameters via a standard computer that does not need to run a real-time operating system.

The real-time process is written in the ADbasic dialect required by the ADwin hardware system.
This dialect provides a fairly high-level interface to the data acquisition and output cards.
It is cross-compiled on a PC, resulting in an ARM binary that can run on the real-time core of the ADwin hardware.
The core task of the real-time process is the continuous evaluation of Eq.~\eqref{eq:df2} for each second-order filter section.
Running at a fixed sampling rate of \samplefreq{}, the platform supports \simulchans{} concurrently running loops, each with \numsos{} individually configurable second-order sections for a total of \totalsos{} sections.

Each sampling interval starts by reading in the current voltage levels at all input channels simultaneously.
For each channel, a settable offset and gain are applied to the converted input values before the second-order sections are calculated.
Each channel's inputs and outputs can be disabled if not used.
Finally, the analog-to-digital converters are pre-populated with the resulting values and scheduled for automatic conversion at the beginning of each sampling interval.
This way, timing jitter because of varying calculation time during each iteration is minimized.
For one output channel at a time, the real-time process can also provide a triangular ramp with user-selectable frequency and amplitude.
Several bit-flags control the status of each loop, such as whether specific second-order sections and inputs/outputs are enabled.
These bit-flags, together with filter coefficients and ramp settings, can be set via the network interface.
In addition, a subset of input and output channels are streamed via the network connection, for monitoring and recording.
The traces are stored in a shared FiFo-buffer (first-in-first-out) which is accessible by the computer to read the data.

On the computer-side, our \nqontrol\ Python package connects to the ADwin system and provides a high-level interface.
It acts as an object oriented library with a simple structure:
The \texttt{ServoDevice} represents one ADwin device containing \simulchans\ \texttt{Servo} objects.
Those servo objects correspond to a specific channel on the physical device and take care of the communication and monitoring with the real-time code.
To simplify the creation of complex filter designs, each \texttt{Servo} contains a \texttt{ServoDesign} object that is implemented in our library \texttt{OpenQlab}\cite{openqlab}.
It provides several convenience functions for creating sequences of filters with up to second order in pole-zero representation, such as integrators, differentiators, lowpass filters and notch filters.
It will show a Bode plot representation of the filters' combined transfer function and can apply the filter to a (measured) transfer function of the plant, i.e.\ the system that is to be controlled.
This allows for a quick iteration in optimizing a set of filters for the individual control task.
We implemented the filter design part in continuous Fourier space, as this is the representation that is probably most common and familiar to physicists.
The filters are automatically transformed into their discrete form before being uploaded as second-order section coefficients into the real-time code.

For example, programming the first control loop to act as an integrator with a corner frequency of \SI{5}{kHz}, then enabling the output and producing a triangular ramp with \SI{30}{Hz} frequency is achieved with the following code sample.
It will also open a plotting window on the computer, which will show a live update of the voltages appearing at the inputs and outputs.
\begin{lstlisting}[language=Python]
from nqontrol import ServoDevice
device = ServoDevice(1)
s = device.servo(1)
s.servoDesign.integrator(5e3)
s.applyServoDesign()
s.outputSw = True
s.enableRamp(frequency=30)
s.realtimePlot() # running in a subprocess
                 # to prevent blocking the
                 # command line
s.disableRamp()
\end{lstlisting}

On top of the Python interface, we have created a web-based, responsive GUI using the Dash framework \cite{dash}, providing an even higher-level interaction with the real-time control system.
Through this GUI, no programming knowledge is required to use the control platform, further lowering the entry barrier to digital control in physics experiments.
Both \nqontrol\ and \texttt{OpenQlab} make heavy use of the Python libraries \texttt{numpy}\cite{numpy}, \texttt{Pandas}\cite{pandas} and \texttt{SciPy}\cite{scipy}.
 \section{Performance}
\label{sec:performance}

    Our digital feedback control system should have a comparable performance to an analog implementation to be an adequate replacement.
    Important performance characteristics of a control system are robustness, noise suppression and recovery time from an external disturbance.
    To evaluate these properties on a realistic example for feedback control in quantum optics, we have set up a test system for locking an optical resonator on a transmission maximum with sub-nanometer precision.
    This test system was then controlled with the digital control implementation presented here, and additionally with a conventional control loop employing analog electronics (operational amplifiers and discrete components) based on a design that has been in use in our group for many years.

\subsection{System characterization}

    \begin{figure}[b]
        \includegraphics[width=\linewidth]{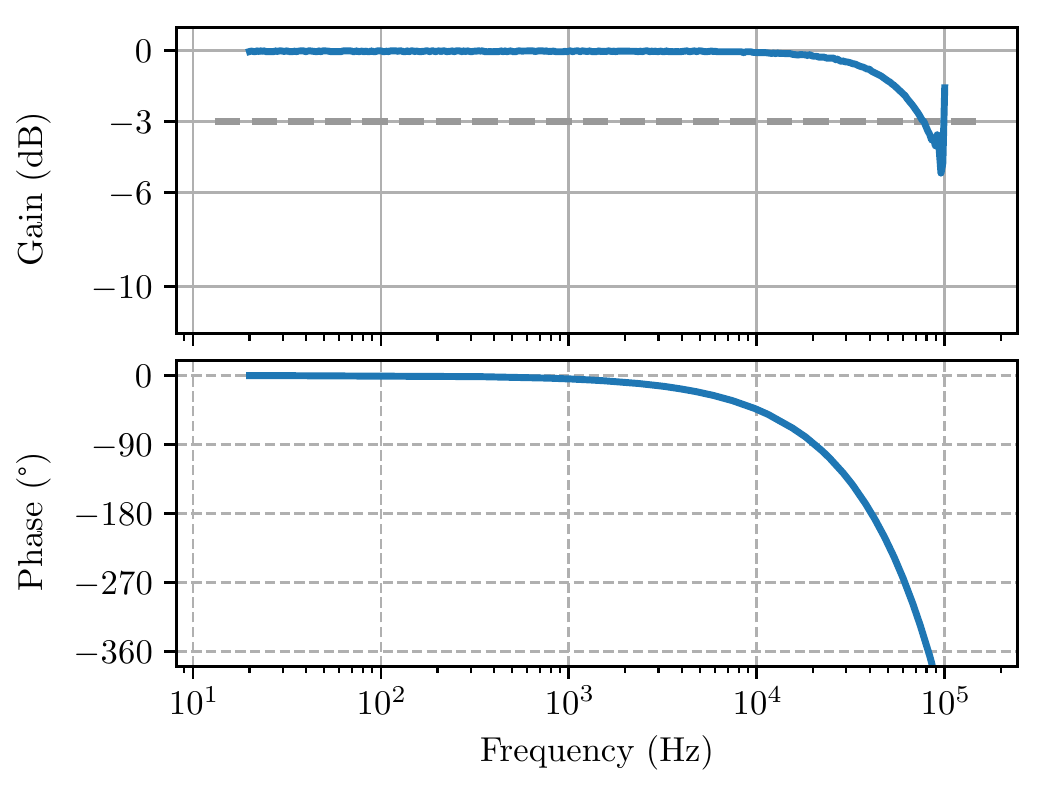}
        \caption{
            Transfer function (Bode plot) of the real-time system running at \samplefreq, for a gain setting of 1. From \SI{10}{kHz} onwards, the phase delay exceeds \SI{45}{\degree}.
        }
        \label{fig:adwin_fra}
    \end{figure}

    To determine the usable bandwidth of our control system, we have measured its transfer function for a unity gain configuration (Figure~\ref{fig:adwin_fra}). A significant phase lag of \SI{45}{\degree} is accumulated at a frequency of \SI{10}{kHz}, while the amplitude stays flat until shortly before the Nyquist frequency (\SI{100}{kHz}), with a \SI{-3}{dB} point at around \SI{80}{kHz}.

    As stated in Section \ref{sec:software}, to keep timing jitter to a minimum, our real-time code always uses a full computing cycle of $1/\samplefreq = \SI{5}{\micro s}$ for the filter calculations. A phase lag of \SI{45}{\degree} at \SI{10}{kHz} corresponds to a time delay of \SI{12.5}{\micro s}, thus another \SI{7.5}{\micro s} of delay were added by the hardware conversion processes.

    In the same unity-gain configuration, we have measured an output noise level of the system of $\SI{480}{nV/\sqrt{Hz}}$ when the analog input was left open. The digital-to-analog conversion on its own showed a noise level of $\SI{260}{nV/\sqrt{Hz}}$.

\subsection{Test system}

    \begin{figure}[t]
        \includegraphics[width=\linewidth]{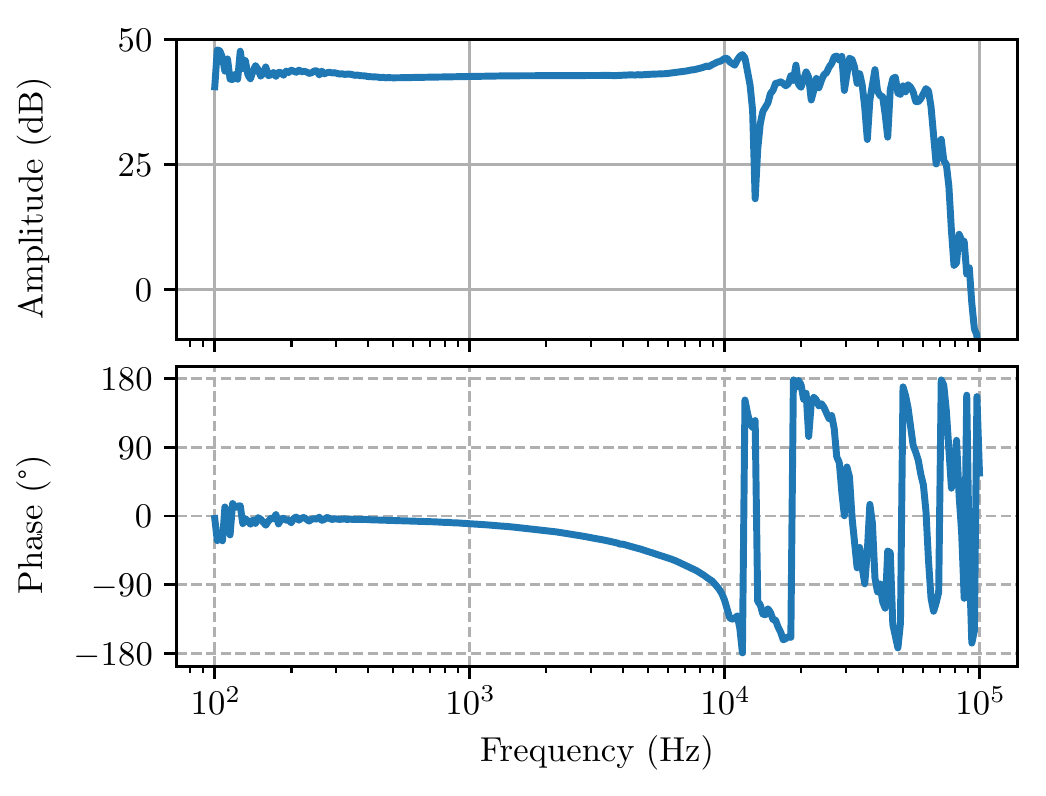}
        \caption{
            Bode plot showing the transfer function for our test system.
            Below $\SI{10}{kHz}$ the response is rather flat, but at higher frequencies the system possesses dispersion-shaped mechanical resonances. 
            Their associated phase delay makes it practically impossible to achieve stable control beyond \SI{10}{kHz} and
            limits the achievable control bandwidth to several kHz.
        }
        \label{fig:mc_fra}
    \end{figure}

    We used a triangular optical ring-cavity \cite{Uehara1997a} for the performance tests as it is a typical system necessary to length-stabilize with sub-micrometer accuracy.
    The cavity had a round-trip length of \SI{42}{cm} and a finesse of about 1000, leading to a FWHM linewidth of roughly \SI{700}{kHz}.
    One of the cavity mirrors was mounted on a piezo actuator which could be driven with $0 \dots \SI{30}{V}$ for precise adjustment of the round-trip length. An error signal for keeping the cavity on resonance was obtained via the Pound-Drever-Hall (PDH) method \cite{Drever1983a}.
    Figure~\ref{fig:mc_fra} shows the measured transfer function of the cavity setup itself. Since this transfer function can only be measured when the cavity is already held on resonance, an initial unoptimised lock of the system with \nqontrol{} was obtained by trial and error. Then, a swept-sine signal was added onto the piezo actuator drive voltage and its response function to the system's error signal was measured. Dividing this response function by the combined drive voltage results in the desired transfer function of just the cavity system by itself.

    \begin{figure}[t]
        \includegraphics[width=\linewidth]{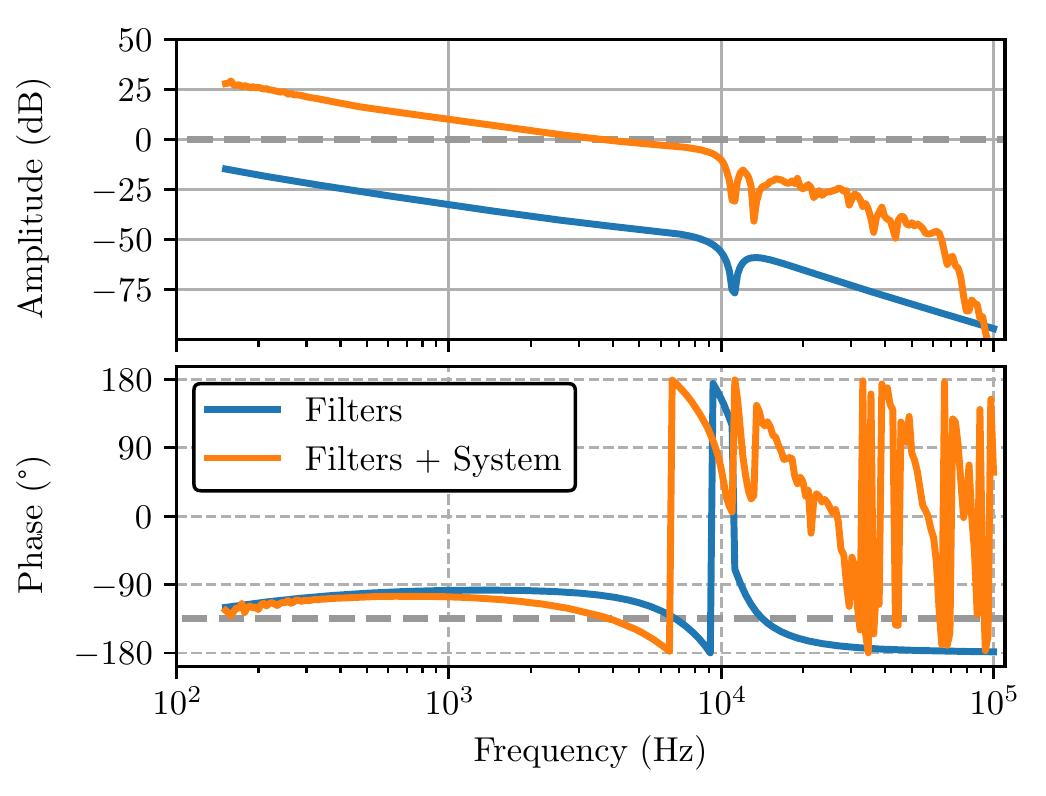}
        \caption{
            Bode plot of the designed filters as given in Table~\ref{tab:servodesign} for the control of our test system (blue).
            The red curve shows the calculated combinated of these filters together with the transfer function of the test system (see fig~\ref{fig:mc_fra}), i.e.\ the overall open-loop transfer function of the control system. 
            This design achieves a \SI{4.5}{kHz} control bandwidth (unity-gain crossing) with a phase margin of \SI{45}{\degree}. 
            The unity-gain level and \SI{-135}{\degree} phase delay are indicated by the dashed grey lines.
        }
        \label{fig:servodesign}
    \end{figure}

    \begin{table}[b]
        \caption{Filter design values used for controlling our optical cavity test system, with corner frequencies and quality factors, where applicable.}
        \begin{tabular}{ll}
        \toprule
            implementation & filters \\
        \midrule
            digital & integrator $\SI{100}{Hz}$ \\
                    & integrator $\SI{10}{kHz}$ \\
                    & 2nd-order lowpass $\SI{9}{kHz}, Q=1$\\
                    & 2nd-order notch $\SI{11.1}{kHz}$, $Q=1$\\\midrule
            analog  & integrator $\SI{100}{Hz}$ \\
                    & integrator $\SI{4}{kHz}$ \\
                    & 2nd-order lowpass $\SI{9}{kHz}, Q=.707$\\
                    & 2nd-order notch $\SI{11}{kHz}$, $Q=1.5$\\
        \bottomrule
        \end{tabular}
        \label{tab:servodesign}
    \end{table}

    Using the measured transfer function, we designed a combination of control filters that together provide high gain at low frequencies and cross the unity-gain point (\SI{0}{dB}) with a phase margin of more than \SI{45}{\degree} to the phase delay of \SI{180}{\degree} which would lead to an amplification of disturbances.
    The unity-gain frequency should be as high as possible and at higher frequencies the gain should stay well below \SI 0 {dB} to avoid an unstable, oscillating system.
    Using those base assumptions, the filter design was tested on the real cavity and has been further optimized for low amplitude noise behind the cavity.
    The resulting filter design and a combination with the system response can be seen in figure~\ref{fig:servodesign} and the values in table~\ref{tab:servodesign}.

    Because of the additional phase delay from the digital feedback loop, the chosen filter values were not the same for the analog and digital implementation, but optimized for each case.

\subsection{Comparing implementation performance}

    \begin{figure}[t]
        \includegraphics[width=\linewidth]{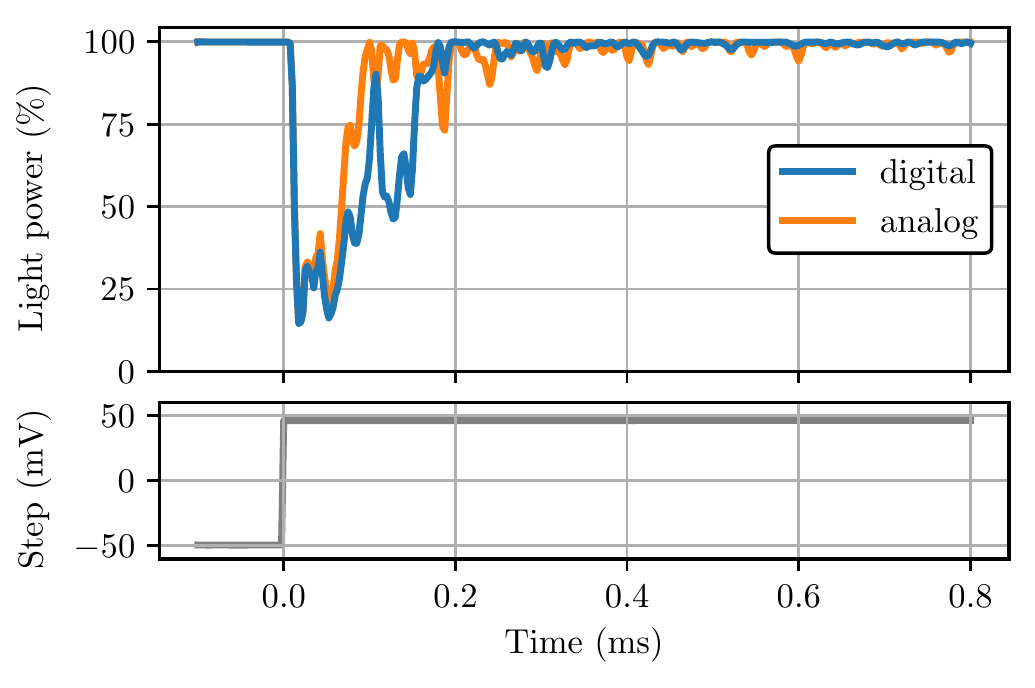}
        \caption{
            Step response of digital and analog control implementation to a \SI{100}{mV} step in piezo voltage occurring at $t=\SI{0}{s}$. 
            Both systems settled back to nominal transmitted power after less than \SI{0.5}{ms}.
            }
        \label{fig:step_answer}
    \end{figure}
    \begin{figure}[b]
        \includegraphics[width=\linewidth]{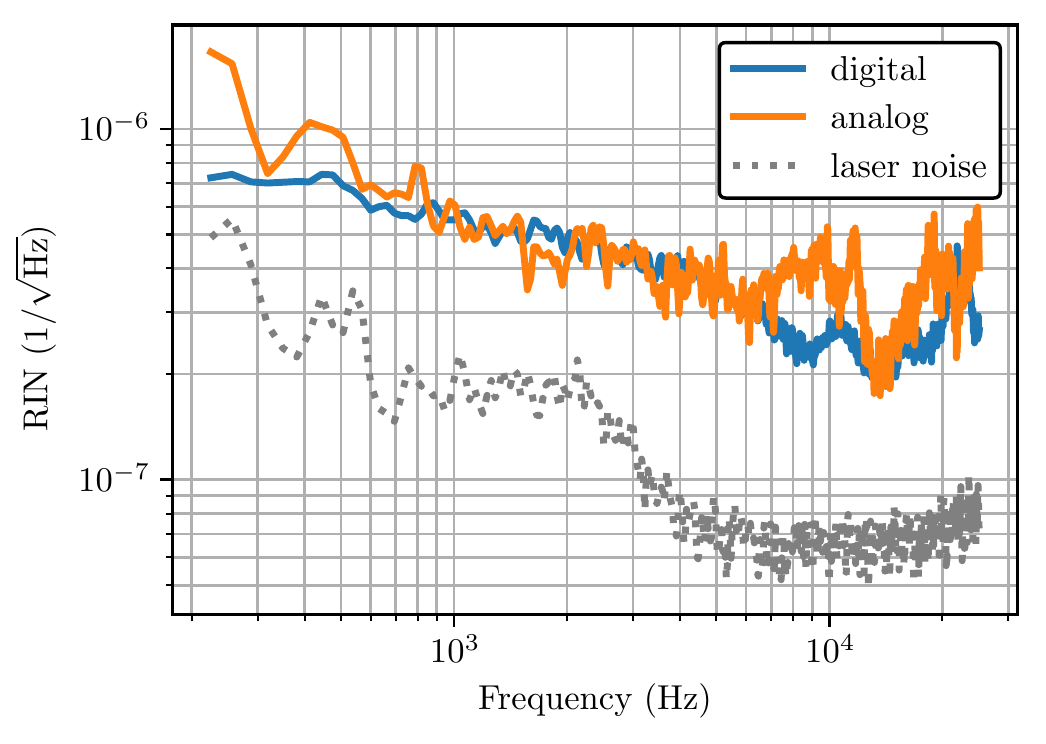}
        \caption{Measurement comparing the relative intensity noise (RIN) levels of digital and analog implementation for our test setup. For all measurements, the same light power was detected ($U_\mathrm{pd}=\SI{6.13}{V}$, equivalent to about \SI{1.4}{mW}). At the measurement frequencies, the laser was not shot-noise limited. Photodiode dark noise was at least one order of magnitude below the measured values and thus not subtracted.}
        \label{fig:servo_noise}
    \end{figure}

    A control system can be characterized by its response to an external disturbance, e.g.\ by a step-like change in one of the system parameters.
    The time it takes for the control system to reach the set point again, and whether it is prone to overshooting and oscillating around the set point then determines the quality of the feedback.
    In our cavity test setup, we implemented such a step response by adding a \SI{100}{mV} step onto the voltage to the piezo element.
    Measuring the power in transmission of the cavity, this caused a drop down to about \SI{20}{\percent} of the transmission compared to the value on resonance.
    The response to this step for both the analog and digital implementation is depicted in \figref{step_answer}. Both controllers were able to cope with the disturbance and the system settled again within less than \SI{0.5}{ms}.

    Another key characteristic of a good control system is noise on the controlled quantity, i.e.\ how well external disturbances are reduced, and how little additional noise is introduced by the control and sensing system itself.
    In our case, a good (out-of-loop) noise figure was the amplitude noise on the transmitted light through the cavity, expressed as relative intensity noise $\mathrm{RIN} = \Delta P/P$.
    For an ideal, noiseless control loop, this amplitude noise would have equaled the amplitude noise on the laser light before it entered the cavity.
    This is the baseline measurement indicated in \figref{servo_noise} as \emph{laser noise}.
    Environmental noise (acoustic noise and cross-coupling of laser phase noise), control noise (from electronics and the piezo element) and sensing noise (from the PDH photodiode) added onto this baseline, resulting in the noise measurement after the cavity.
    We compared the noise level obtained with our conventional analog control circuit with the digital system and obtained similar results.
    At frequencies below around \SI{5}{kHz}, both control implementations were most likely limited by sensing noise, as evidenced by the fact that further increasing the gain actually increased the noise level.
    Above \SI{5}{kHz}, the digital control implementation was less noisy.
    This might be explained by a slightly detuned notch filter in the analog implementation, because of component tolerances and their temperature drift.
    Here, the flexibility and quick turn-around time of filter adjustments in the digital control system came to full strength.
    More importantly, however, we were able to show that the digital control system did not introduce additional noise from the analog-digital-analog conversion steps and is therefore a suitable replacement for analog controllers in the feedback control tasks of our experiments in quantum optics.
 \section{Conclusion and Outlook}
\label{sec:outlook}

We have developed and tested an open-source control platform, based on commercially available hardware and with the aim of providing a flexible, high-performance control solution for experiments in quantum optics.
Supporting \simulchans{} simultaneously active control loops with a sampling rate of \samplefreq{}, we believe our solution can cover a wide range of control tasks.
Building on high-quality, long-term supported hardware components, we demonstrated that our system can reach comparable performance to more conventional, analog circuitry.
Both real-time code and the interface code has been made available as open source, such that interested parties can adapt the system to their needs and integrate it into existing lab environments and control infrastructures.
We actively encourage participation and code contribution to further maintain and advance the system.
 
\begin{acknowledgments}
This research has been funded by the Deutsche Forschungsgemeinschaft (DFG, German Research Foundation) -- 388405737.\\
This article has LIGO document number P1900343.
\end{acknowledgments}

\bibliography{references}

\end{document}